\documentclass{jpsj3}

\title{Magnetic Phase Diagram of Alternating Chain Compound $\mathrm{Pb_2V_3O_9}$}

\author{Kazuhiro NAWA$^{1}$\thanks{E-mail address: knawa@kuchem.kyoto-u.ac.jp}, Chishiro MICHIOKA$^{1}$, Kazuyoshi YOSHIMURA$^{1}$
\thanks{E-mail address: kyhv@kuchem.kyoto-u.ac.jp}, \\
Akira MATSUO$^{2}$, and Koichi KINDO$^{2}$%\\
}
\inst{
$^{1}$Department of Chemistry, Graduate School of Science, Kyoto University, Kyoto 606-8502, Japan
$^{2}$Institute of Solid State Physics, University of Tokyo, Kashiwanoha, Kashiwa, Chiba 277-8581, Japan
}
\abst{$\mathrm{Pb_2V_3O_9}$ is a quasi one-dimensional compound
with antiferromagnetic alternating chains.
We measured the magnetization and the specific heat in a single crystal of $\mathrm{Pb_2V_3O_9}$
with applied magnetic fields up to 14 T and established the magnetic phase diagram.
The phase boundary in the high magnetic field region
is in a good agreement with that predicted by the mean field theory in the classical magnetic ordering of the transverse magnetization,
while that in the low magnetic field region should be discussed in the viewpoint of the Bose-Einstein condensation of the field induced magnons.
Thus, the experimentally determined phase diagram shows the asymmetry, which should be originated in the mixing contribution of excited states as well as in the interchain interactions.} 
\kword{$\mathrm{Pb_2V_3O_9}$, Bose-Einstein condensation, magnons, alternating chain, field induced magnetic ordering, magnetic phase diagram}

\begin{document}
\maketitle

\section{Introduction} 
\begin{figure}
\centering
\includegraphics[width=6cm]{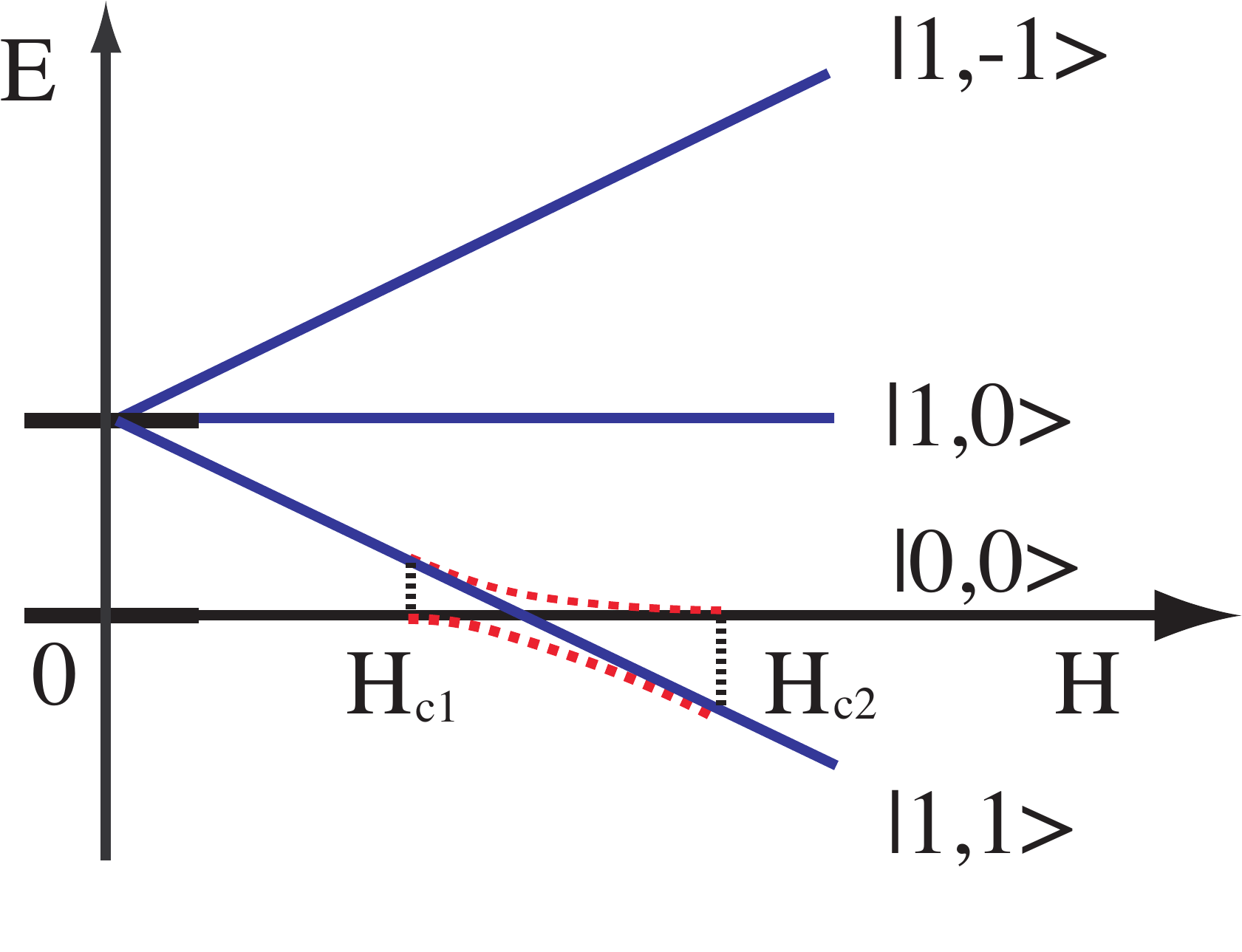}
\caption{(Color online)
Energy levels of the spin states of $| S, S_z \rangle$ = $| 0, 0 \rangle$, $| 1, 1 \rangle$, $| 1, 0 \rangle$ and $| 1, -1 \rangle$ in the dimers without any interdimer interactions (solid lines)
and the mixed $| S, S_z \rangle$ = $| 0, 0 \rangle$ and $| 1, 1 \rangle$ states in interacting dimers derived by the mean field theory (dashed lines).\cite{MFT}}
\label{zeeman}
\end{figure}

Low dimensionality in quantum spin systems has been a matter of particular interest over some decades 
since it could lead to novel exotic phenomena.
%Low-dimensional compounds have been in interest since they would have strong quantum fluctuations and cause exotic properties.
Since the ground state in the low-dimensional system has instabilities with quantum fluctuations,
it can be controlled by external fields or impurity effects.
%In the system with strong quantum fluctuations and the close energy levels of Fermi-liquid state and long-range ordered state,
%its ground state could change by a magnetic field, a mechanical pressure, or a chemical impurity substitution.
%Spin gap compound is one of such systems.
A weakly coupled spin gap system is a candidate for exhibiting novel quantum phenomena due to its low dimensionality.
%From viewpoint of such quantum phase transitions, a field induced magnetic phase transition in a spin gap system is in large attention with interest.
In the case of dimer systems without any interdimer interactions, 
%Spin gap system consists of spin dimers coupled by antiferromagnetic interactions. Each
a dimer has the ground state of $|S,S_z \rangle = |0,0 \rangle$ with excited triplet states of
$|S,S_z \rangle = |1,1 \rangle, |1,0 \rangle, |1,-1 \rangle$.
The field dependence of the energy levels is shown in Fig. \ref{zeeman} as solid lines.
The external magnetic field splits the degenerated triplet states 
and reduces the energy gap between the $|0,0 \rangle$ and $|1,1 \rangle$ states. 
%Above the field which is required to close the energy gap,
Under the magnetic field, the ground state changes from the $|0,0 \rangle$ to $|1,1 \rangle$ state and spins induced by the field are aligned along the field.
%When a magnetic field is applied and the spin gap is closed, the ground state of no interacting dimers shows crossover from the product of the singlets
%to the state in which all spins are aligned by the field.
In the presence of the additional antiferromagnetic interdimer interactions,  
the field-induced magnetic moments can exhibit the magnetic ordering of their transverse spin components.
%In contrast, in the case of a system with strong interdimer interactions,
%an antiferromagnetic long range ordering (AFLRO) of the spin components perpendicular to the field occurs 
The antiferromagnetic long range ordering (AFLRO) in this manner was predicted by the mean field theory\cite{MFT}.
The AFLRO state was approximated by the ground state of mixed $|0,0 \rangle$ and $|1,1 \rangle$ states.
The energy levels of the mixed states derived based on the mean-field magnetic ordering model of the transverse magnetization
are shown in Fig. 1 by dashed lines.
In addition, the symmetric bell-shape phase boundary can be derived\cite{MFT}.
The lower critical field $H_{c1}$, the upper critical field $H_{c2}$ and the maximum transition temperature to AFLRO phase $T_\mathrm{max}$
can be calculated from the intradimer and interdimer interactions.
At low temperatures, the phase boundary becomes almost perpendicular to $H$ axis. 
%Additionally, further studies were made in the compounds %
%The phase transition to AFLRO state has been studied so far and explained from a mean field theory\cite{MFT}
% and a description of Bose-Einstein Condensation (BEC) of magnons\cite{TlCuCl3}. Previous studies
In fact, $\mathrm{TlCuCl_3}$\cite{TlCuCl3,TlCuCl3_2},
$\mathrm{KCuCl_3}$\cite{KCuCl3},
%$\mathrm{BaCuSi_2O_6}$\cite{BaCuSi2O6}, 
$\mathrm{Ba_3Cr_2O_8}$\cite{Ba3Cr2O8},
$\mathrm{Sr_3Cr_2O_8}$\cite{Sr3Cr2O8}
and $\mathrm{Pb_2V_3O_9}$\cite{Pb2V3O9,Pb2V3O9_2,Pb2V3O9_3,Pb2V3O9_4}
%Previous studies in these compounds indicate that AFLRO state is described better by a Bose-Einstein condensation (BEC) of magnons in terms of
%the two experimental features.
were found to show the field-induced magnetic orderings.
The induced $|1,1 \rangle$ excitations would have characteristics of bosons as excited magnons.
The AFLRO phase can be interpreted in terms of a Bose-Einstein condensation (BEC) of magnons.
%Some characteristics of the ordering , the upturn of the magnetic susceptivity, 
%should be explained by the Bose-Einstein condensation (BEC) of magnons.
%One is the magnetic susceptivity which shows the minimum behavior in the temperature dependence.
The increase of the magnetic susceptivility below the transition temperature is explained by the increase of condensed 
magnon density. %due to the condensation of magnons.
%because temperature-independent susceptivity is predicted in the classical transverse ordering.
%The other is the phase boundaries which is described by the following power low
The magnon BEC picture also gives a power-law like critical behavior described as
\begin{equation}
\label{powerlaw}
T_c \propto [H_{c1} - H_{c1}(T=0)]^{1/\phi} ,
\end{equation}
where $T_c$ is the transition temperature to AFLRO phase and $\phi$ is the critical exponent. 
%In contrast, the mean field theory predicted a nearly perpendicular line to $H$ axis at low temperatures.
Most experimental data supported the magnon BEC picture. %with some differences. %but also included some inconsistencies. 
%Most experimental data support the magnon BEC theory, the data also show some features which cannot be explained.
%The first one is slight disagreement in the value of $\phi$.
The critical exponent $\phi$ was experimentally reported to be between 1.5 and 2.3, while the value was theoretically predicted as 1.5\cite{TlCuCl3}.
%in the magnon BEC picture
%The other is asymmetries in magnetic phase diagrams.
%The magnetic phase boundaries predicted from the mean field theory is symmetric for the midpoint of the critical fields $H = (H_{c1} + H_{c2})/2$
%because of the particle - hole symmetry.
%Such diagrams are observed experimentally in $\mathrm{Sr_3Cr_2O_8}$\cite{Sr3Cr2O8}, $\mathrm{Pb_2V_3O_9}$\cite{Pb2V3O9_2}.
%From these viewpoints, we made further investigation on the phase diagram by studying alternating chain compound $\mathrm{Pb_2V_3O_9}$.
Furthermore, from the view of the phase diagram topology, 
asymmetric $H - T$ phase diagrams was reported in $\mathrm{Sr_3Cr_2O_8}$\cite{Sr3Cr2O8} and $\mathrm{Pb_2V_3O_9}$\cite{Pb2V3O9_2},
while symmetric one was expected because of the particle - hole symmetry even in the case with the BEC model.

\begin{figure}
\centering
\includegraphics[width=8cm,bb=0 0 974 715]{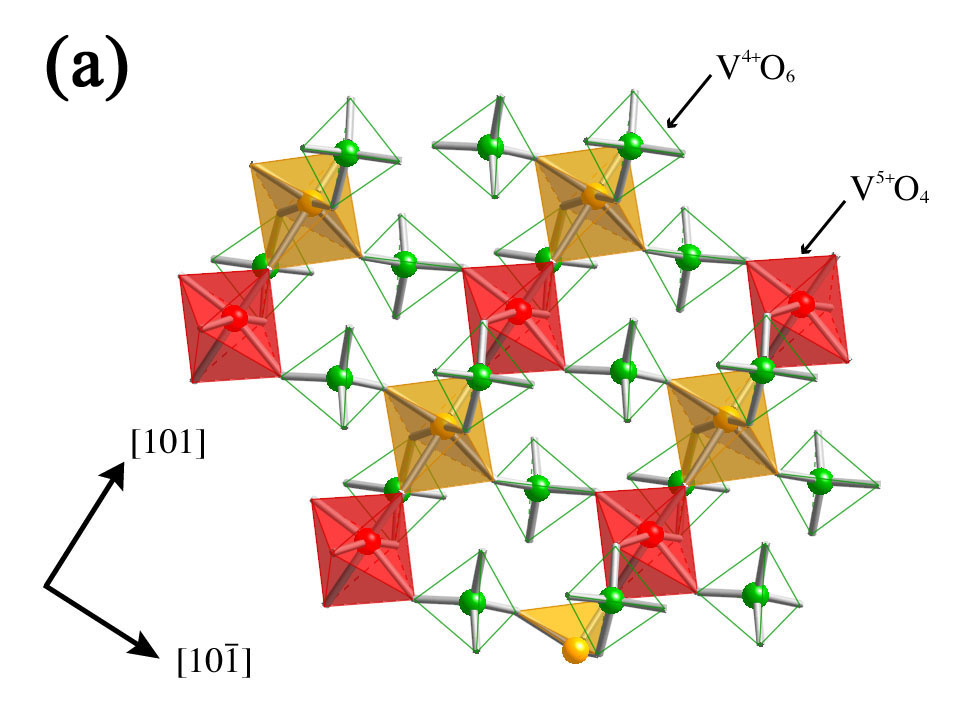} \\
\includegraphics[width=7cm]{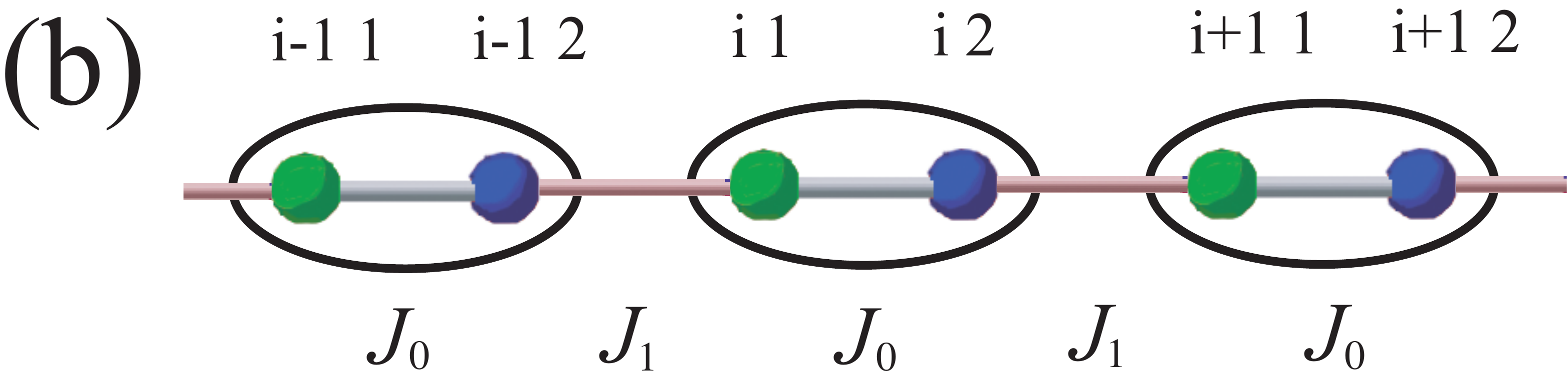}
\caption{(Color online)
(a) The arrangement of $\mathrm{V^{4+}O_6}$ octahedrons and $\mathrm{V^{5+}O_4}$ tetrahedrons in the $ac$ plane.
(b) The illustration of the alternating chain with the notations above and the interaction parameters below the chain, 
both of which are used in eq.~(\ref{Pb_H})(see text).}
\label{structure}
\end{figure}

$\mathrm{Pb_2V_3O_9}$ was first synthesized by the electrochemical reduction of molten $\mathrm{PbV_2O_6}$\cite{Mentre1} and
found to have antiferromagnetic alternating chains.
The arrangement of the chains in $\mathrm{Pb_2V_3O_9}$ is shown in Fig. \ref{structure}(a) and the alternating chain is illustrated in Fig. \ref{structure}(b).
A magnetic $\mathrm{V^{4+}}$ ions with coordinating six oxygen atoms form $\mathrm{V^{4+}O_6}$ octahedrons
with corner-sharing of oxygen atoms, forming the spin chain with S = 1/2 along the [101] direction.
The octahedrons are aligned on a tilt and the distances between adjacent $\mathrm{V^{4+}}$ ions change alternatively.
Along the [10$\mathrm{\overline{1}}$] direction, $\mathrm{V^{5+}O_4}$ tetrahedrons stand between $\mathrm{V^{4+}O_6}$ octahedrons.
The calculation by the quantitative spin dimer analysis based on tight binding calculations suggested that 
the $\mathrm{V^{4+}}$-O-O-$\mathrm{V^{4+}}$ super-superexchange interactions 
along the [10$\mathrm{\overline{1}}$] direction were dominant for the antiferromagnetic alternating chain model\cite{Mentre2}.
%The crystal structure and the magnetic properties of $\mathrm{Pb_2V_3O_9}$ are explained in terms of an antiferromagnetic alternating chain.
From magnetization measurements, the existence of a spin gap and the phase transition to the AFLRO phase under magnetic fields have been clarified\cite{Pb2V3O9}.
Lower and upper critical fields have been determined as $H_{c1} = 4$ T and $H_{c2} = 38$ T, respectively.
From the power low behavior of phase boundary,
the critical exponent has been determined as $\phi = 1.9$\cite{Pb2V3O9}, or $2.0$\cite{Pb2V3O9_3,Pb2V3O9_4}.
%The previous papers are focused on critical behavior and the behavior of the whole asymmetric phase diagram is less discussed.
In this paper, we report the magnetic phase diagram precisely determined by using a single crystal in order to elucidate the asymmetric $H - T$ phase diagram.
%and discuss the magnetic phase diagram in terms of interdimer interactions.

\section{Experiments}
The polycrystalline sample of $\mathrm{Pb_2V_3O_9}$ was prepared by solid state reaction from PbO,
$\mathrm{V_2O_5}$ and $\mathrm{V_2O_3}$.
The powders were mixed and then sintered at 570 $\mathrm{C}^\circ$ for 2 days with intermediate grindings.
Single crystals of $\mathrm{Pb_2V_3O_9}$ were grown by self-flux and floating zone methods.
The crystal growth by self-flux method was made by cooling the sintered sample slowly from 750 $\mathrm{C}^\circ$ in an evacuated quartz tube.
Glossy crystals were obtained with a maximum size of 0.2 $\times$ 0.5 $\times$ 0.2 $\mathrm{mm^3}$.
By floating zone method, 8 g of powdered sample was molded into a rod under pressure of 400 $\mathrm{kgf/cm^2}$.
The rod was sintered at 500 $\mathrm{C}^\circ$ for two days in vacuum
and then was set to the sample holder under Ar atmosphere of 1 atm.
The crystal growth was made by melting the sample rod partially 
with a crystal growth speed of 0.5 mm/h. 
The size of obtained glossy crystal is about 2 $\times$ 2 $\times$ 1 $\mathrm{mm^3}$.
A Debye-Scherrer X-ray diffraction measurement was performed to check the purity,
and the single crystal X-ray diffraction measurement was done by transmission Laue method to determine the crystal plane, and axes.
%Both measurements were made at room temperature.
%To examine the magnetic properties, the magnetic susceptibility, the high field magnetization curve, and the specific heat under field were measured.
The magnetic susceptibility was measured by a superconducting quantum interference device magnetrometer.
The magnetization curve was measured up to 55 T with a pulse magnet at Ultra High Magnetic Field Laboratory, Institute for Solid State Physics, 
University of Tokyo. 
The specific heat was measured by relaxation method under magnetic field up to 14 T.

\section{Results}
\begin{figure}
\centering
\includegraphics[width=4cm,bb=0 0 678 521]{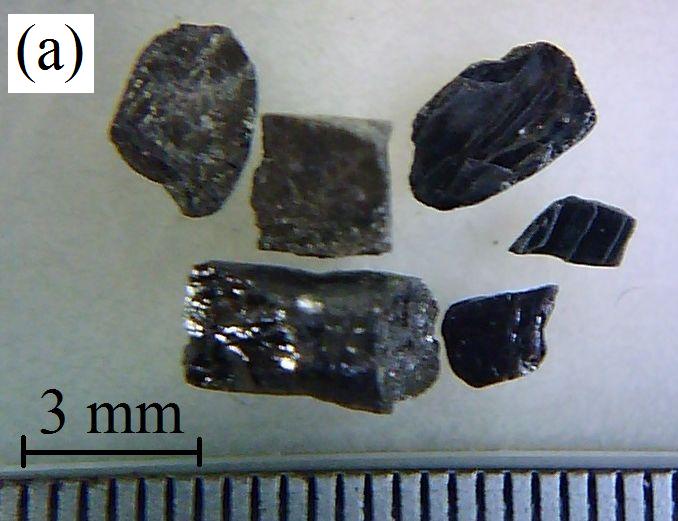} \\
\includegraphics[width=4cm,bb=0 0 1404 1126]{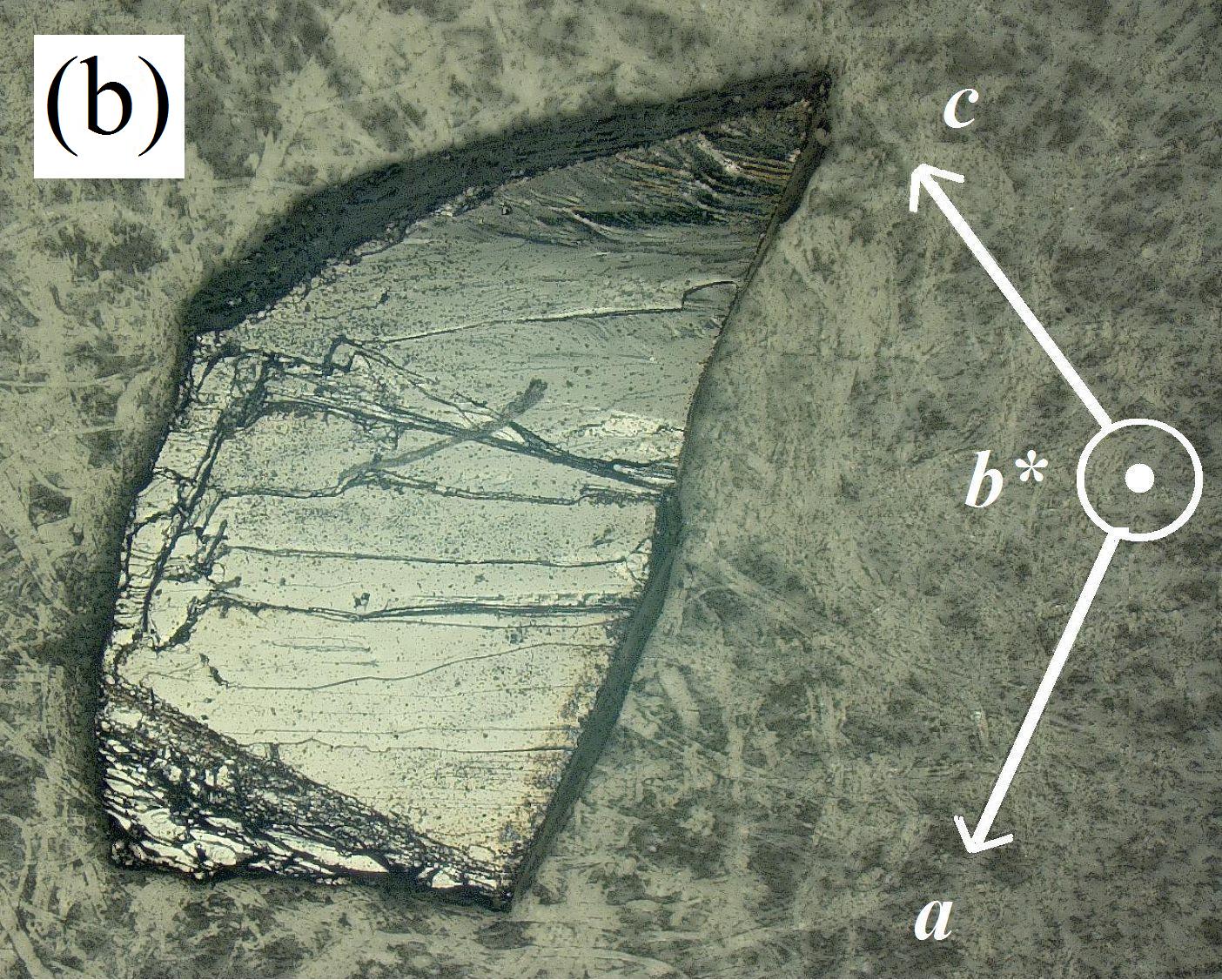}
\caption{(Color online)
(a) Viewgraphs of single crystals obtained by floating zone method.
(b) Crystal axes of the single crystal determined by transmission Laue method.}
\label{crys}
\end{figure}

\begin{figure}
\centering
\includegraphics[width=8cm,clip]{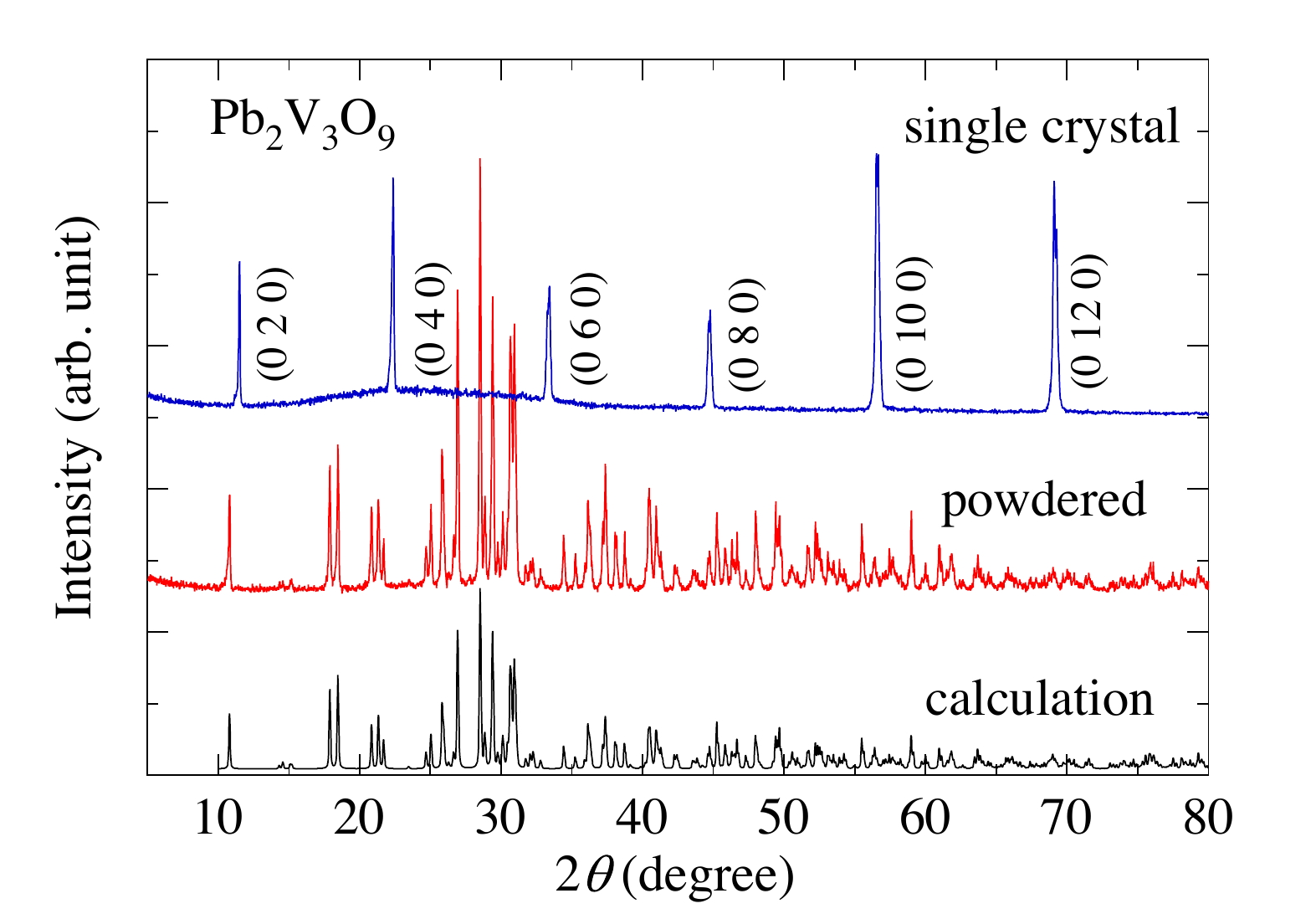}
\caption{(Color online) 
Debye-Scherrer X-ray diffraction patterns for the single and powdered crystals with calculated powder pattern for comparison.}
\label{Xray}
\end{figure}

Viewgraphs of single crystals grown by floating zone method are shown in Fig. \ref{crys}(a).
%The $b^*$ plane was obtained in the samples.
Figure \ref{Xray} shows the Debye-Scherrer X-ray diffraction patterns.
The crashed powders of single crystals were confirmed to be in a single phase.
In addition, each single crystal was found to have a single domain by transmission Laue X-ray photographs.
The crystal axes of the sample used in measurements of physical properties were determined as shown in Fig. \ref{crys}(b).

\begin{figure}
\centering
\includegraphics[width=8cm,clip]{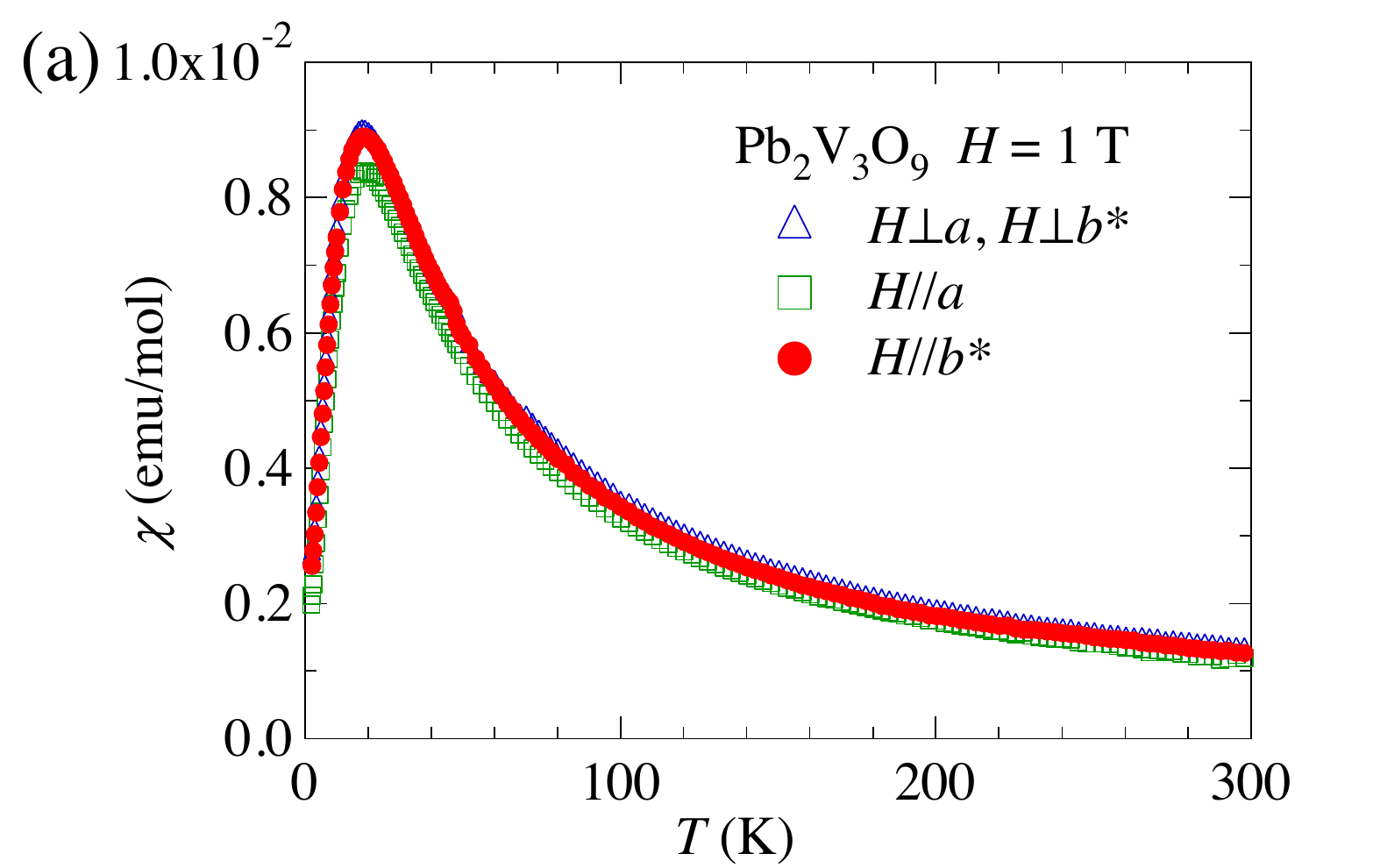} \\
\includegraphics[width=8cm,clip]{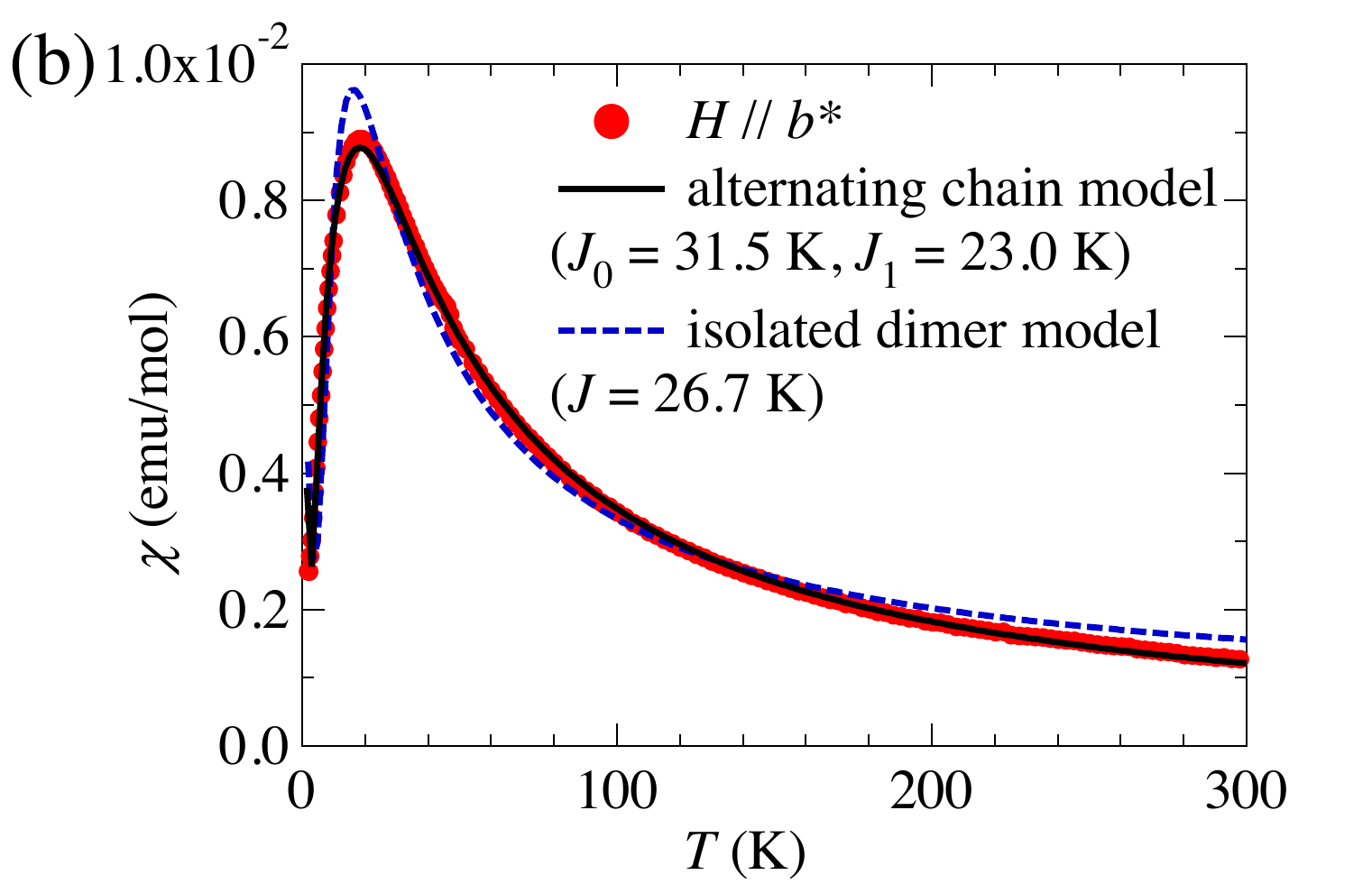}
\caption{(Color online)
(a) Temperature dependence of magnetic susceptibility under the applied field of 1 T parallel and perpendicular to $a$, $b^*$ axes.
(b) Temperature dependence of magnetic susceptibility with $H \parallel b^*$ (closed circles).
Dashed and solid lines show the fitted lines
with the isolated dimer\cite{dimer} and the alternating chain \cite{Johnston} models, respectively.}
\label{susceptibility}
\end{figure}

Figure \ref{susceptibility}(a) shows the temperature dependence of the magnetic susceptibility under the magnetic field of 1 T.
The susceptibility does not depend on the field direction so much.
This fact suggests that the magnetic anisotropy is not so large that the orbital contribution would be small.
To estimate spin interactions, the susceptibility was fitted to the isolated spin dimer model\cite{dimer}
as well as to the antiferromagnetic alternating chain model\cite{Johnston}.% as shown in Fig. \ref{susceptibility}.
The fitting functions are described by
\begin{equation}
\chi_0 = \chi_\mathrm{spin}(g, J_0, J_1) + \frac{C}{T} + \chi_0 ,
\end{equation}
where $J_0, J_1, g, C, \chi_0$ are the intradimer interaction,
the interdimer interaction, g-factor, the Curie constant and the temperature independent term, respectively.
The values of estimated parameters are shown in Table \ref{parameters}.
In the case of the isolated dimer, of course, $J_1$ must be zero.
%Difference in the values of estimated interactions $J_0, J_1$ due to the field direction was 0.6 K at the largest.
%From comparing the experimental data in Fig. \ref{susceptibility}(b) with the two fitting curves,
%From comparison of the two fitting curves with the experimental curve in Fig. \ref{susceptibility}(b),
The alternating chain model reproduces the experimental data much better than the isolated dimer model does,
indicating the presence of the interdimer interaction.
Here, the $J_0$ and $J_1$ were determined as 31.5 and 23.0 K, respectively.
%Magnetic field dependence in $J_0, J_1$ were estimated within 0.6 K at the largest.
%The system of $\mathrm{Pb_2V_3O_9}$ is found to have strong interdimer interactions between the dimers from the estimation.

\begin{figure}
\centering
\includegraphics[width=8cm,clip]{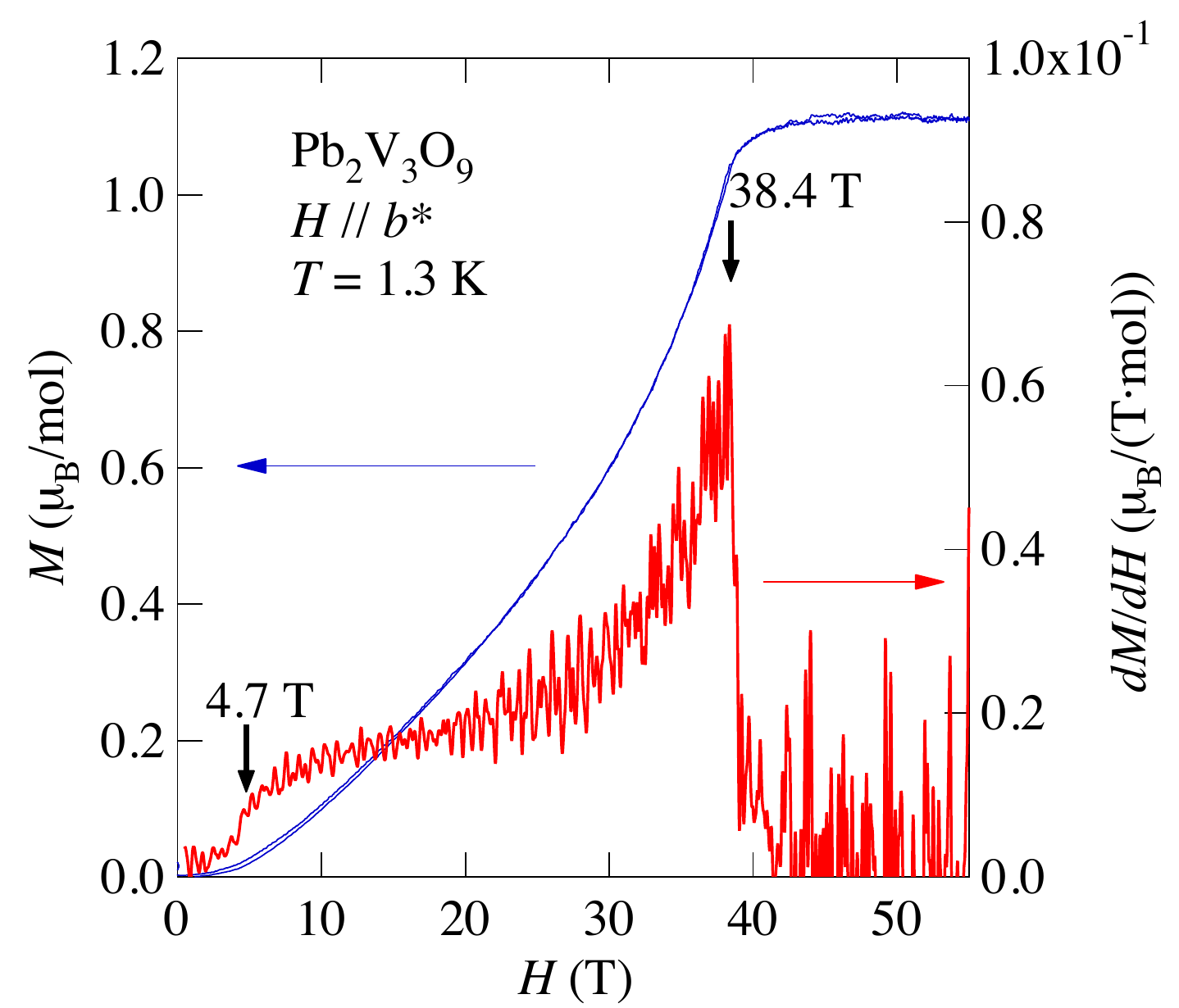}
\caption{(Color online)
Field dependence of the magnetization with the field $H \parallel b^*$.
Critical field is determined as an anomaly field of first derivative in magnetization curves.}
\label{magnetization}
\end{figure}

The high-field magnetization %measurement was made by using a single crystal to determine the upper critical field $H_{c2}$. The magnetization
curve measured at 1.3 K is shown in Fig. \ref{magnetization}.
The whole behavior is almost consistent with that observed in the previous report using the polycrystalline sample\cite{Pb2V3O9_2}.
The critical fields in 1.3 K were determined as $H_{c1} = 4.7$ T and $H_{c2} = 38.4$ T from anomalies in the field derivative of the magnetization curve.
%The estimated critical fields are close to that in powdered sample.
%close to the value in the previous measurement ($T = 1.52$ K, $H_{c1} = 5.2$ T, $H_{c2} = 36.8$ T).
%The small differences in the critical fields are attributed to the weak anisotropy.

\begin{figure}
\centering
\includegraphics[width=8cm,clip]{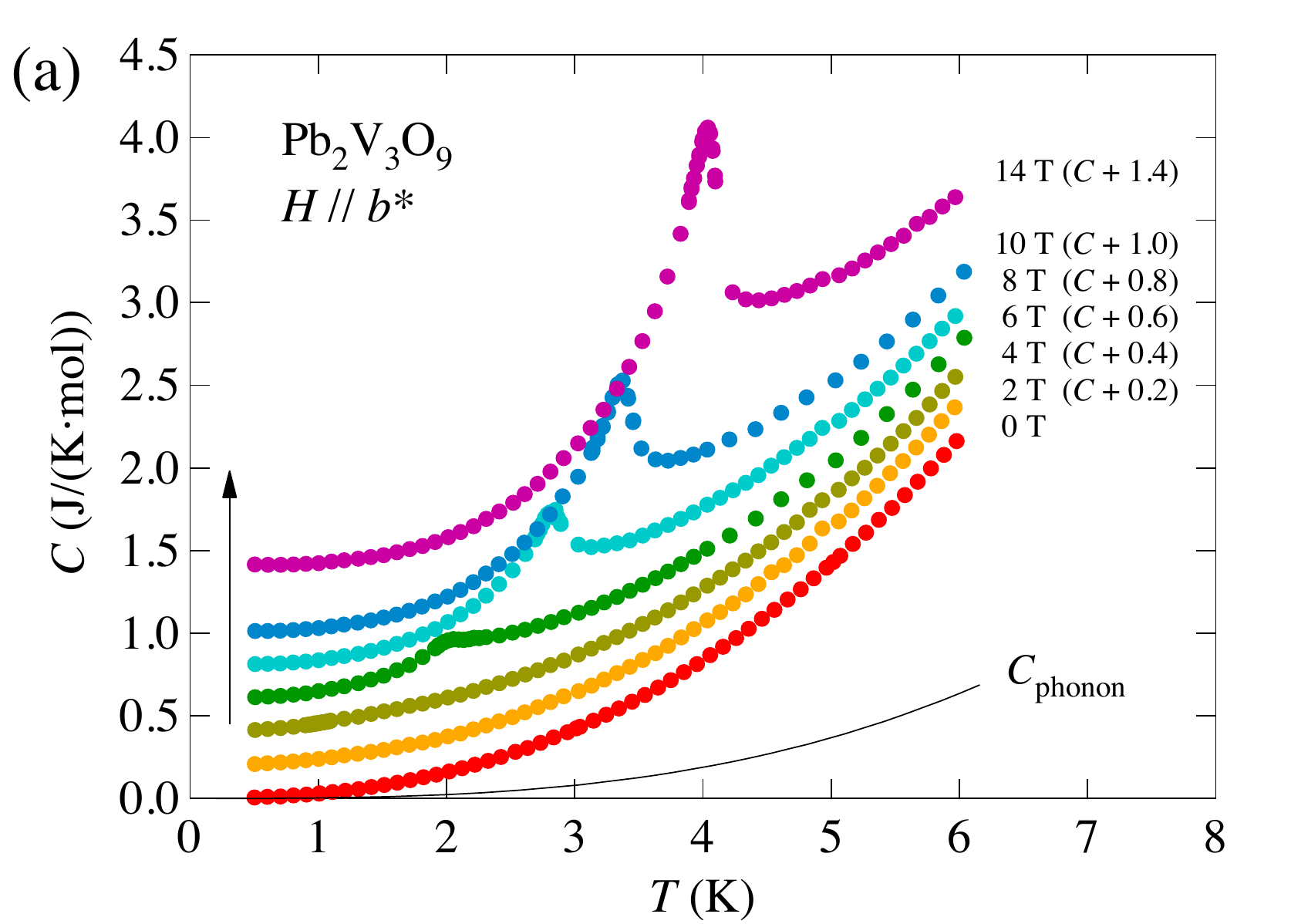} \\
\includegraphics[width=8cm,clip]{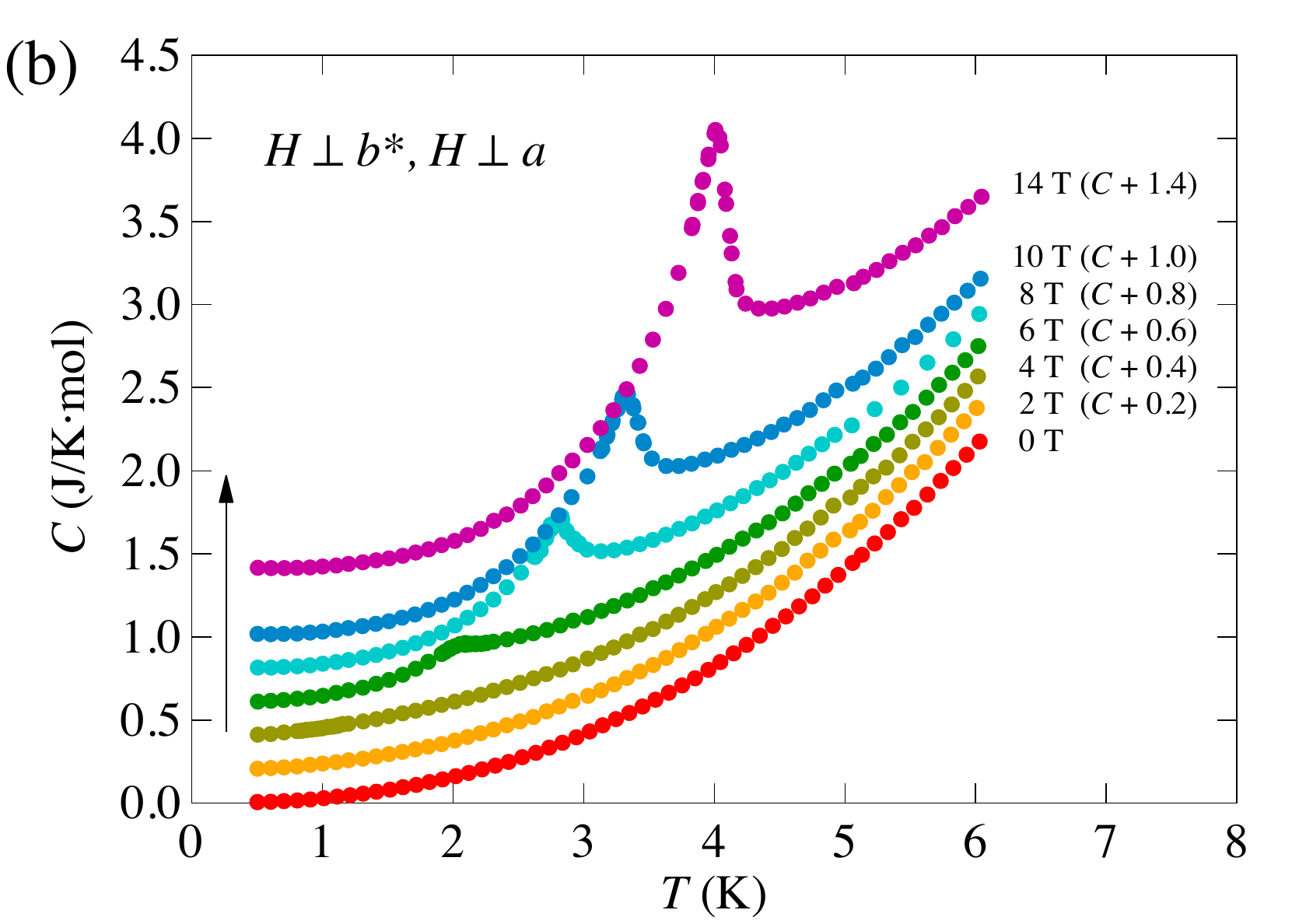}
\caption{(Color online) 
Temperature dependences of specific heats with the field (a) $H \parallel b^*$ and (b) $H \perp b^*$,
with offset of $H$ $\times$ 0.1 J/(K $\cdot$ mol). Solid line shows the phonon contribution.}
\label{HC}
\end{figure}

\begin{figure}
\centering
\includegraphics[width=8cm,clip]{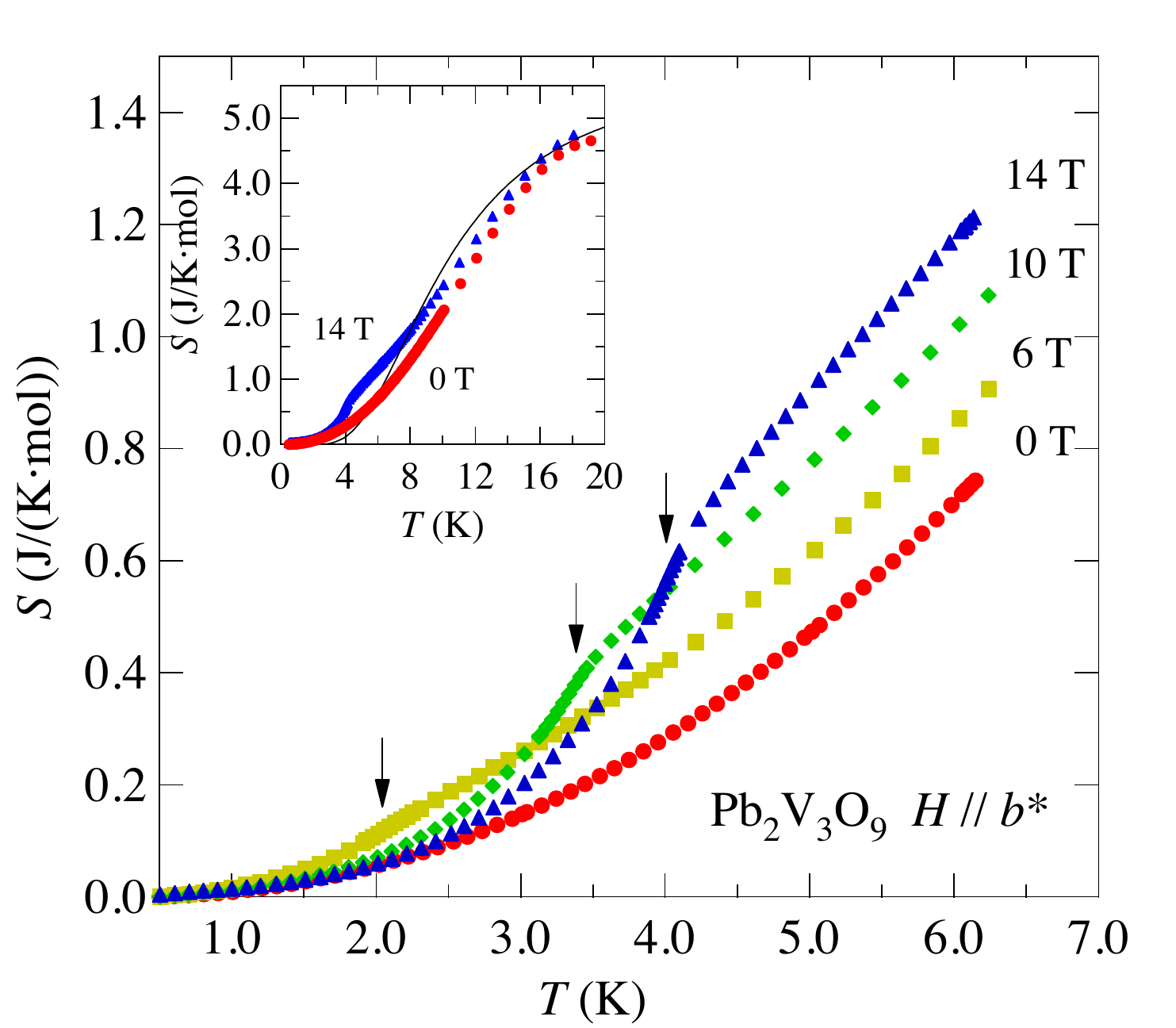}
\caption{(Color online) 
Temperature dependence of magnetic entropy with the field $H \parallel b^*$.
Solid arrow indicates $T_c$ at each field.
Solid line in inset shows the magnetic entropy of the isolated dimer with the spin gap $\Delta = 26.7$ K under zero field.
%Calculation was made by integrating $C_\mathrm{mag}/T$ to exclude contribution of phonons.
}
\label{entropy}
\end{figure}

The specific heat was measured to determine the phase diagram precisely.
Figure \ref{HC} shows the specific heat at temperatures between 0.5 and 6 K with the field applied along or perpendicular to the $b^*$ axis.
Below 4.5 K, a $\lambda$-type anomaly due to the second order phase transition was observed, which corresponds to the field induced magnetic ordering. %to AFLRO phase.
%The value of the transition temperature $T_c$ was not affected by field directions.
%This fact indicates that interactions between spins are thought to be isotropic Heisenberg type.
With increasing field, the peak shifts to the higher temperature and the intensity is enhanced.
The shift of peak by the magnetic field can be explained by the description of BEC of magnons.
The critical temperature $T_c$ was determined from the second temperature-derivative of 
$\{ C(H) - C(H = 0) \}/T$,
where $C(H)$ is the specific heat under the field $H$.
%The difference in strength of the anomalies depended on the entropy 
%change accompanied by the phase transition.
%The jump at the peak temperature enhances with increasing the field, indicating the enhancement of the entropy change.
%which was yielded in the transition from AFLRO phase to paramagnetic phase. 
%We also estimates the entropy change in the magnetic entropy around the transition temperature.
%The fact that no Schottky peak was observed below 20 K indicates
%that the system cannot be described by isolated dimers.
%the magnetic specific heat is much smaller than the phonon contribution.
%In contrast, the fact that the value of the specific heat at 6 $\sim$ 18 K became smaller with the increasing field suggests
%that the magnetic specific heat is as large as the phonon contribution 
%because the field dependence of Schottky-type specific heat becomes very weak in the higher temperature.
%Since it was difficult to separate the magnetic specific heat from the phonon contribution,
Rough estimation of the entropy changes accompanied by the phase transition was given by integrating $C(H)/T$
after extracting a phonon contribution of $C_\mathrm{phonon} \sim 2.94 \ T^3$ \ mJ/(K $\mathrm{\cdot}$ mol) from $C(H)$, 
as shown by solid line in Fig. \ref{HC}.
The phonon estimation was made with an assumption that the entropy change in the alternationg chain model agrees with
that in the isolated dimer model at high temperatures.
The $J_0$ in Table \ref{parameters} was used for the entropy estimation of the isolated dimers.
Figure \ref{entropy} shows the estimated entropy with the field applied along the $b^*$ axis.
%We roughly excluded the phonon contribution from the entropy by integrating $C_\mathrm{mag}/T$,
%hypothesizing that the phonon contribution is independent on the field strengths.
%The magnetic entropy, $\Delta S = S(H) - S(H = 0)$, is shown in Fig. \ref{entropy}.
%The field-direction of the entropy is very small.
The entropy estimation with the field perpendicular to $b^*$ axis provides a quite similar result.
%The entropy in the field over 6 T becomes smaller above 7 K.
%This behavior results from the subtraction of the entropy in 0 T.
%inversion between the value of $C(H)$ and $C(H = 0)$.
%At 20 K, the entropy becomes 4.6 J/(K $\mathrm{\cdot}$ mol), which is 80 \% of the total magnetic entropy 
%expected for 1 mol of 1/2 spins (R $\ln$ 2).
%The value of the entropy should converge to 0 in the high temperature limit, since the value in each field, even in zero field, should converge to $R \ln 2 = 5.76$ J/(K mol).
%The entropy for isolated dimers shows a close value, which ensures the correct estimation of the phonon contribution.
The entropy is increased with increasing the field and suppressed below $T_c$.
The enhancement by the field results from the reduction of the spin gap 
and the suppression below $T_c$ indicates the magnetic ordering under the magnetic field. 
%The entropy enhancement $S(H) - S(H = 0) $ under the field of 14 T becomes 0.46 J/(K mol).
In the field of 6 T, the entropy is estimated as 0.12 J/(K $\cdot$ mol), which corresponds to the 2.1 \% of the maximum value $R \ln 2 = 5.76$ J/(K mol)
expected for the spin system of S = 1/2.
Therefore, the critical magnon density in 6 T ($T_c = 2.1$) can be described as $n_c$ = 0.021
and becomes about 10 times larger than that in $\mathrm{TlCuCl_3}$\cite{TlCuCl3_2} at the same temerature.
This indicates that more magnons are induced in $\mathrm{Pb_2V_3O_9}$ than in $\mathrm{TlCuCl_3}$.
%Above 16 K, the entropy in 2 $\sim$ 14 T are almost same.
%The magnetic susceptibility shows the maximum at this temperature and the short range order almost disappears.
%The close values of entropy under a field indicate that they reached the maximum value $R \ln 2 = 5.76$ J/(K mol).
%In spite of the enough magnetic entropy, the existence of strong interdimer couplings makes the separation difficult
%between the magnetic specific heat and the phonon contribution.
%Since separation between the magnetic specific heat and the phonon contribution was difficult in spite of the enough magnetic entropy,
%we can expect complex mechanisms such as a coupling between the interdimer interactions and the lattice deformation.

The phase diagram was determined as shown in Fig. \ref{diagram1}.
Whole behavior of the phase boundary does not depend on the field direction.
The anisotropic interactions could yield the field-dependence of the phase diagram, and
% and the phase diagram was asymmetric for the midpoint of $H_{c1}$ and $H_{c2}$
%The isotropic behavior indicates that the ordered state is not an Ising antiferromagnetic state 
%but the state controlled by Heisenberg interactions.
%Generally, the dependence on the field direction results from single ion anisotropy and 
%anisotropic interactions such as Dzyaloshinskii-Moriya interactions.
%The presence of Dzyaloshinskii-Moriya interactions is expected in the spin Hamiltonian
%since $\mathrm{Pb_2V_3O_9}$ crystal does not have inversion symmetry.
in $\mathrm{Pb_2V_3O_9}$, e.g., the Dzyaloshinskii-Moriya interaction could work due to the absence of the inversion symmetry
at the center of $\mathrm{V^{4+}}$ pairs.
%The independence on the field direction indicates that those anisotropic interactions are weak and
In the case of $\mathrm{Pb_2V_3O_9}$, however, the weak field-direction dependencies in $\chi$, $C$ and the phase diagram of $\mathrm{Pb_2V_3O_9}$ indicate that 
the anisotropic interactions are small.
%These facts are also supported by magnetic susceptibility, which shows little dependence on field directions.

\section{Discussion}
\begin{figure}
\centering
\includegraphics[width=8cm,clip]{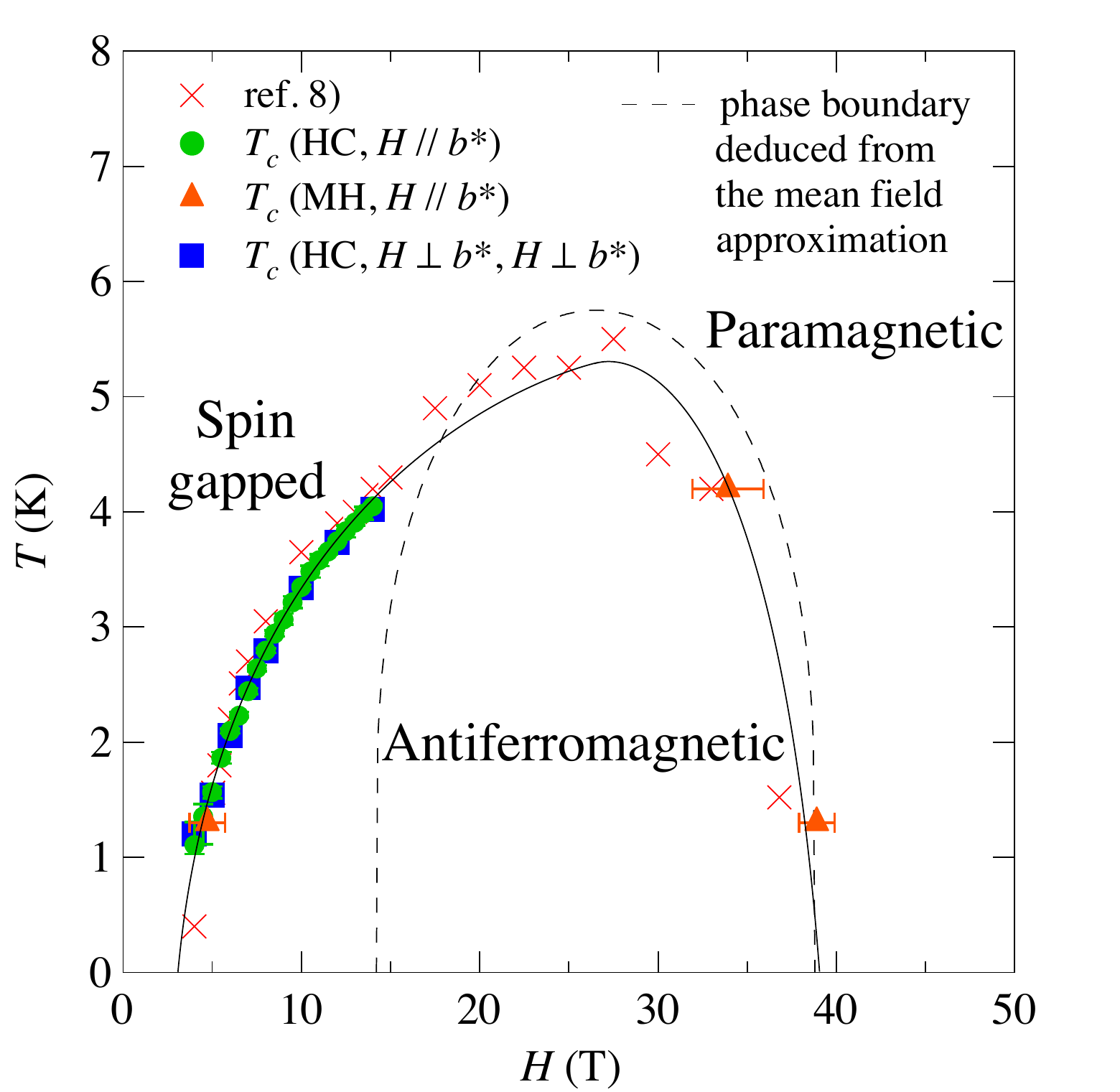}
\caption{(Color online)
The $H - T$ magnetic phase diagram of $\mathrm{Pb_2V_3O_9}$.
Solid line is a guide for eyes. Dashed line corresponds to deduced phase boundaries from classical mean field approximation\cite{MFT} based on 
the parameters in Table \ref{parameters}.}
\label{diagram1}
\end{figure}

Here we discuss the shape of the phase diagram and its asymmetric behavior in this section.
The dashed line in Fig. \ref{diagram1} expresses the phase boundary of the classical transverse magnetic ordered phase,
which can be produced by the mean field theory with the interaction parameters estimated from the $\chi - T$ curve.
%By use of the estimated values $J_0$ = 31.5 K, $J_1$ = 23.0 K, $g$ = 2.09,
%Considering the crystal structure and the magnetic behavior, interactions are thought to arrange one dimensionally.
%Though exchange paths are expected to be whether along [101] direction or along [10-1] direction,
%the direction of the interactions have no influence on the expected diagram.
The spin Hamiltonian of one-dimensional alternating chain is described by 
\begin{equation}
	\begin{split}
H &= H_0 + H' \\
H_0 &= J_0 \sum_i \mathbf{s}_{i1} \cdot \mathrm{s}_{i2} - g \mu_B \mathrm{H} \cdot \sum_i ( \mathbf{s}_{i1} + \mathbf{s}_{i2} ) \\
H' &= J_1 \sum_{i} \mathbf{s}_{i2} \cdot \mathbf{s}_{i+1 \ 1} ,
 %+ J_2 \sum_{<i,j>_2} \mathbf{s}_{i1} \cdot \mathbf{s}_{j2} + J_3 \sum_{<i,j>_3} \mathbf{s}_{i2} \cdot \mathbf{s}_{j1}
	\end{split}
\label{Pb_H}
\end{equation}
where $i$ represents the notation of the dimer and $i1$ and $i2$ denote the two spins on each dimer.
The site i and i+1 locate on the nearest neighbor along the spin chain.
The arrangement of the interaction parameters and the notations are shown in Fig. \ref{structure}(b).
The direction of the spin chain is not important in the following discussion.
According to the mean field theory, Tachiki and Yamada confined themselves to subspace of the $|1,1\rangle$ and $|0,0\rangle$ states in the vicinity of $H_{c1}$.\cite{MFT}
The model Hamiltonian in eq.~(\ref{Pb_H}) is rewritten into the following low energy effective Hamiltonian as,
\begin{equation}
	\begin{split}
H &= H_0 + H' \\
&= \frac{N}{2} \left( J_0 + \frac{J_1}{8} - g \mu_{B} H \right) \mathbf{1} - \frac{1}{2} \left( J_0 + \frac{J_1}{4} - g \mu_{B} H \right) \sum_i \sigma_{iz} \\
& \ \ \ + \frac{J_1}{16} \sum_{i} \large[ -2(\sigma_{ix} \sigma_{i+1 \ x} + \sigma_{iy} \sigma_{i+1 \ y}) + \sigma_{iz} \sigma_{jz} \large] .
	\end{split}
\label{effPb_H}
\end{equation}
From eq.~(\ref{effPb_H}), $H_{c1}$, $H_{c2}$ and $T_\mathrm{max}$ can be estimated as
\begin{equation}
	\begin{split}
H_{c1} &= \frac{1}{g\mu_B} \left( J_0 - \frac{J_1}{2} \right) = 14.2 \ \mathrm{T} \\
H_{c2} &= \frac{1}{g\mu_B} \left( J_0 + J_1 \right) = 38.8 \ \mathrm{T} \\
T_\mathrm{max} &= \frac{J_1}{4k_B} = 5.8 \ \mathrm{K}.
\label{estH}
	\end{split}
\end{equation}
%The deduced phase boundaries are shown in Fig. \ref{diagram1} as a dashed curve.
The estimated value of $H_{c2}$ is quite close to the experimental one.
The magnetic field at which $T_{c}$ becomes the highest is estimated to be $H = (H_{c1}+H_{c2})/2 = 26.5$ T,
which is also quite close to the experimental value as shown in Fig. \ref{diagram1}.
In contrast, the estimated $H_{c1}$ deviates largely from the experimental value.
%The value of $H_{c1}$ can be affected by interchain interactions, while they also change $H_{c2}$ to the same extent. 
%The correspondence in high field region suggests that we should consider not only interchain interactions
%but also another factor which cause asymmetry in the phase boundaries.

\begin{figure}
\centering
\includegraphics[width=8cm,clip]{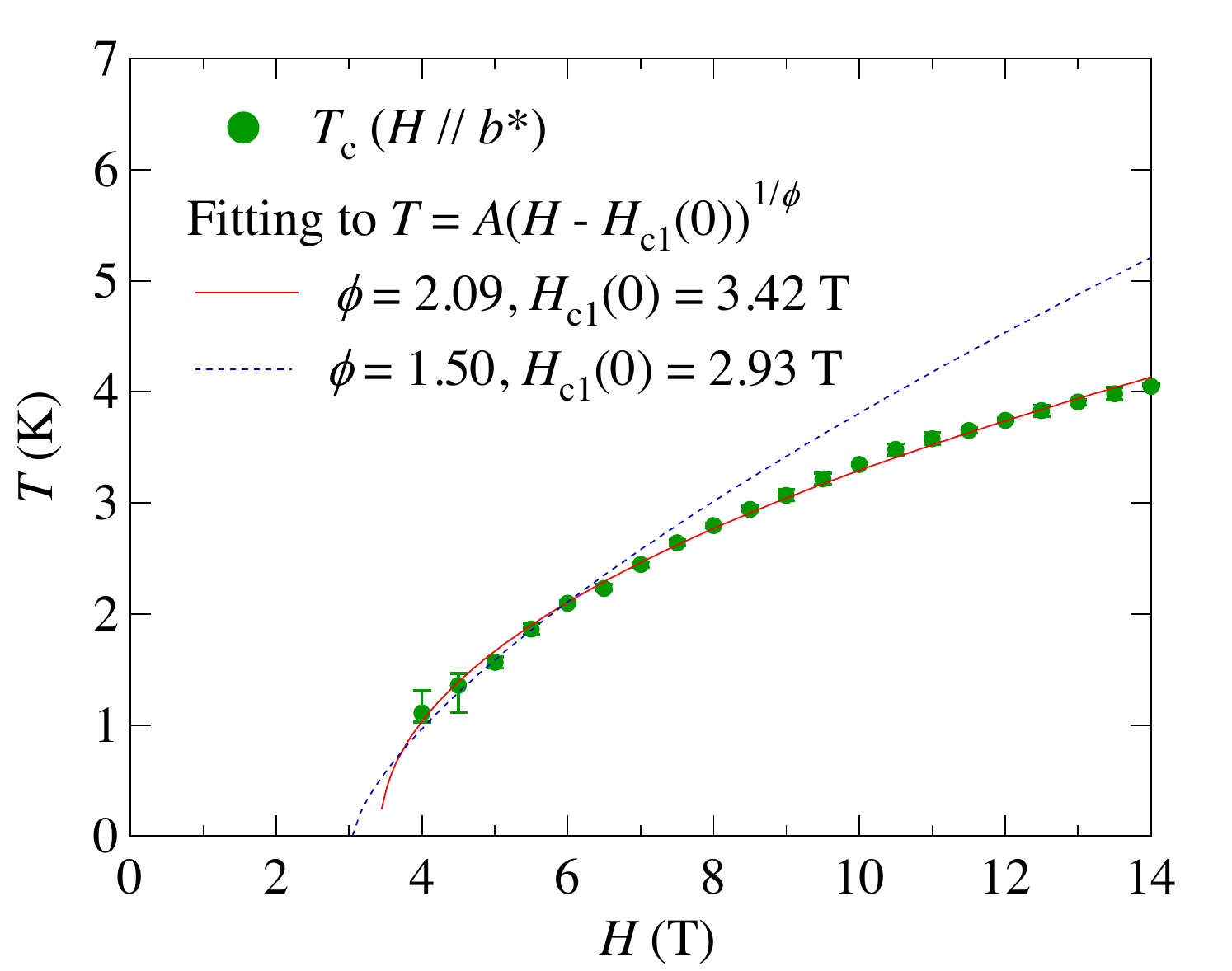} 
\caption{(Color online)
Magnetic phase diagram focused on at low temperatures and in the low field region.
Dashed and solid lines are obtained by power law as $T = A (H - H_{c1}(T=0))^{1/\phi}$ with critical exponent $\phi$ = 1.50
and $\phi$ = 2.09, respectively.
%The data below 6 T were used for the above fitting since the fitting curve cannot explain the whole data. 
%In the case that both critical exponent $\phi$ and lower critical field $H_{c1}$ were used for a fitting to  power law,
%they were determined as $\phi$ = 2.09, $H_{c1}(0)$ = 3.42 T.
%Resulting fitting curve is shown as a solid curve.
}
\label{diagram2}
\end{figure}

Figure \ref{diagram2} shows the phase boundary focused on the lower critical field $H_{c1}$.
The power law behavior of eq.~(\ref{powerlaw}) was predicted in the magnon BEC model.
The dashed and solid lines in Fig. 8 are the phase boundaries produced by the fitting to eq.~(\ref{powerlaw}).
The dashed line is made by fitting the experimental points under 6 K with $\phi$ confined to the value of $1.5$ and
the solid line is made by using $\phi$ as a fitting parameter, which is determined as 2.1.
While the critical exponent $\phi$ is expected to be 1.5 in the magnon BEC model,
the experimental data are better explained by $\phi = 2.1$ with $H_{c1} = 3.4$ T.
%differs from the predicted curve in the mean field theory,
%which is regarded as a straight line perpendicular to the field axis at $H_{c1}$. 
%To examine the critical behavior, the curve fitting is made by the least-square fitting to eq.~(\ref{powerlaw}) with $\phi$ confined to a constant $1.5$.
%The fitting curve deviates from the experimental data.
%When both $\phi$ and $H_{c1}$ are used for the fitting, they are determined as $\phi = 2.1$ and $H_{c1} = 3.4$ T,
In early papers, the critical exponent exhibits $\phi = 1.9$\cite{Pb2V3O9}, or 2.0\cite{Pb2V3O9_3,Pb2V3O9_4} and
the possibility of $\phi$ convergence to the universal value $\phi = 1.5$ was suggested in the low-temperature limit,
which was first observed in the experimental phase boundary of $\mathrm{TlCuCl_3}$.
The BEC theory by using realistic dispersion relations of $\mathrm{TlCuCl_3}$ explained the phase boundary below 5 K
and $\phi = 1.5$ was obtained from fitting in the temperature range below 2.4 K\cite{TlCuCl3_2}.
Also in the case of $\mathrm{Pb_2V_3O_9}$, the valid description of the magnon dispersion should be necessary %in the vicinity of $H_{c1}$.
for the universality of AFLRO phase.
%We fitted the phase boundaries in the temperature range of $4 \leq T \leq 6$ K to a dashed curve, 
%considering the prediction that the the critical exponent may converge to the theoretical value $\phi = 1.5$ in extreme low temperarure\cite{Pb2V3O9_4}.
%we cannot discuss validity of the prediction since the determination of transition temperature becomes more difficult as the temperature approaches zero.
%as long as estimation is made from the eq.~(\ref{effPb_H}).
%As well as the mean field theory, the magnon BEC picture cannot explain the origin of the asymmetry in the phase diagram. 
%For these reasons, we consider that it is important to obtain a description of the whole phase diagram.
%The disagreement is not unique to the mean field approximation.
%We can also obtain the same values of $H_{c1}$ and $H_{c2}$ 
%This is because the BEC picture and the mean field theory describe quite simular situation in the quantum critical point,
%where the magnon density is 0 (at $H_{c1}$) or 1 (at $H_{c2}$).
%The fact that the deduced phase boundaries well correspond to the experimental boundaries only in high field region

To evaluate the magnon-magnon interactions more correctly, we took into consideration the additional effects
of the $| 1,0 \rangle$ and $|1,-1 \rangle$ states.
In the subspace of the $| 0,0 \rangle$ and $| 1,1 \rangle$ states,
a point inversion symmetry $h_\mathrm{eff} \rightarrow  - h_\mathrm{eff}, \sigma \rightarrow  -\sigma$
in Hamiltonian eq.~(\ref{effPb_H}) ensures the symmetry of the phase diagram for a zeropoint of the effective field $h_\mathrm{eff} = h - J_0 - J_1/4$.
%Therefore it is important to discuss asymmetry of the experimental phase boundaries in terms of the break in the inversion symmetry.
%When we simply ignore the high energy states and confine ourselves to the subspace of singlet state $| 0,0 \rangle$ and $|1,1 \rangle$ state as in eq.~(\ref{effPb_H}),
%the relative energy levels of the two states is invariant for the inversion.
%The inversion changes the relative energy levels of $| 0,0 \rangle$, $|1,1 \rangle$, $| 1,0 \rangle$ and $|1,-1 \rangle$.
Second-order processes between the $| 0,0 \rangle$, $| 1,1 \rangle$ states and the excited $| 1,0 \rangle$, $|1,-1 \rangle$ states yield the perturbation terms and
they break the inversion symmetry in the spin Hamiltonian.
%A key to consider the excited states is reported\cite{disptheory1,disptheory2}.
The mixing effect with the excited states was discussed in Refs. 15 and 16.
In these references, a bond-operator technique was used to describe both the spin-liquid and the ordered states
in the field and pressure induced quantum phase transition in $\mathrm{TlCuCl_3}$.
The values of $H_{c1}$ and $H_{c2}$ were described by intradimer and interdimer interactions, and
magnetization curves and the magnon dispersion relations observed in $\mathrm{TlCuCl_3}$ were accounted for by the bond-operator technique.
%The operators are defined as:
%\begin{align}
%s^\dagger |0> &\equiv \frac{1}{\sqrt{2}} (|\uparrow \downarrow > - |\downarrow \uparrow >) \\
%t_+^\dagger |0> &\equiv - |\uparrow \uparrow > \\
%t_0^\dagger |0> &\equiv \frac{1}{\sqrt{2}} (|\uparrow \downarrow > + |\downarrow \uparrow >) \\
%t_-^\dagger |0> &\equiv |\downarrow \downarrow >
%\label{operators}
%\end{align}
%which correspond to the creations of $| 0,0 \rangle$, $|1,1 \rangle$, $| 1,0 \rangle$ and $|1,-1 \rangle$ states, respectively.
In this theory, 
the operators which correspond to creations and annihilations of the $| 0,0 \rangle$, $|1,1 \rangle$, $| 1,0 \rangle$ and $|1,-1 \rangle$ states 
represent spin degrees of freedom on each dimer.
The interdimer interactions yield the mixed ordered states which are approximated by a linear combination of the $|0,0 \rangle$,
$|1,1 \rangle$ and $|1,-1 \rangle$ states.
The $|1,0 \rangle$ state does not contribute its term to the mixed states.
%in the process of transformation to the bond-operator representation.
%Operators of the ground state and the excited triplet states are approximated by a linear combination of dimer singlets and triplets.
%\begin{align}
%a_i &= u s_i + v(f e^{i\mathbf{Q} \cdot \mathbf{r}_i} t_{i+} + g e^{i\mathbf{Q} \cdot \mathbf{r}_i} t_{i-}) \\
%b_{i+} &= u(f t_{i+} + g t_{i-}) - v e^{i\mathbf{Q} \cdot \mathbf{r}_i} s_i \\
%b_{i0} &= t_{i0} \\
%b_{i-} &= f t_{i-} - g t_{i+}
%\label{func} 
%\end{align}
%The wave vector $\mathbf{k} = \mathbf{Q}$ corresponds to the dispersion minimum of the lowest triplet mode
%and $u$, $v$, $f$, and $g$ are field dependent coefficients.
%Although the ground state has the pure triplet character with the values of $u = 0$, $v = 1$, and $f = g = 0$) in the vicinity of $H_{c2}$,
%it has the singlet character specified by $u = 1$, $v = 0$ and finute values of $f$ and $g$ in the vicinity of $H_{c1}$, 
In consideration of the mixing contribution of the $|1,-1 \rangle$ state with the $|0,0 \rangle$ and $|1,1 \rangle$ states,
$H_{c1}$ of $\mathrm{Pb_2V_3O_9}$ is obtained as
\begin{equation}
%	\begin{split}
H_{c1} = \frac{1}{g\mu_B} \left( \sqrt{J_0^2 - J_0 \cdot J_1} \right) = 11.7 \ \mathrm{T} \\
%H_{c2} &= \frac{1}{g\mu_B} \left( J_0 + J_1 \right) = 38.8 \ \mathrm{T} .
%	\end{split}
	\label{estH2}
\end{equation}
and $H_{c2}$ becomes exactly the same as that in eq.~(\ref{estH}).
The $H_{c2}$ in eq.~(\ref{estH}) agrees with the experimental $H_{c2}$.
The $H_{c1}$ in eq.~(\ref{estH2}) is smaller than that in eq.~(\ref{estH}) and becomes closer to the experimental $H_{c1}$.
The asymmetric behaviors especially in the vicinity of $H_{c2}$ and $H_{c1}$ 
are due to the change in mixing contribution of the $|1,-1 \rangle$ state.
Around $H_{c2}$, the mixing contribution is ignorable and therefore
the magnon-magnon interactions in the high field region can be described by quasi one-dimensional mean field.
In the vicinity of $H_{c1}$, the finite mixing contribution reduces the minimum of the magnon dispersion and
%The reduction of the dispersion minimum makes the energy gap smaller between the ground state and the excited state.
makes $H_{c1}$ smaller than the value in eq.~(\ref{estH}).
In addition, magnon-magnon interactions are well described by interacting dilute bosons and 
therefore the phase boundary shows power-law behavior.  
%In the low field region, an energy gap between a singlet state and a $|1,-1 \rangle$ state is small.
%The small energy gap results in strong mixing contribution near $H_{c1}$.
%It is considered that the mixing contribution becomes strong near $H_{c1}$ since an energy gap between a singlet state
%and a single $|1,-1 \rangle$ state is smaller in low field region.
%Actually, the experimental value of $H_{c1}$ is still smaller than the value in eq.~(\ref{estH2}).
The alternating chain model with the field-induced magnetic ordering explains the physical properties of $\mathrm{Pb_2V_3O_9}$ very well. 
Details of AFLRO state in $\mathrm{Pb_2V_3O_9}$, especially the critical exponent, may be explained with the modified alternating chain model 
by additional small interchain interactions.
The experimental $H_{c1}$ is reproduced well if the value of $J_1$ is regarded as 30.5 K.
Since the assumed $J_1$ is not so different from that estimated from $\chi-T$ curve,
the presence of interdimer interactions may increase $J_1$ and also affect the critical exponent.

\section{Conclusion}
In summary, the magnetic phase diagram of $\mathrm{Pb_2V_3O_9}$ was determined precisely from %measurements of
the high field magnetization and the specific heat by using single crystal.
The experimental phase boundary near $H_{c2}$ coincides with that based on the mean field picture,
while that in the vicinity of $H_{c1}$ shows the power-law behavior predicted in the magnon BEC picture.
These results indicate that the mixing contribution of excited states with low-lying states shifts
the phase boundary to the lower field especially in the case of $H_{c1}$.
%Furthermore, asymmetry of $H - T$ phase boundaries is predicted to become stronger with increasing interdimer interactions.

\acknowledgments
We thank Dr. T. Waki for helpful discussion and provision of $\mathrm{V_2O_3}$ samples.
The single crystal X-ray diffraction measurement was carried out using facilities of 
Research Center for Low Temperature and Materials Sciences, Kyoto University.
This work was supported by a Grant-in-Aid for Scientific Research from the Japan Society for Promotion of Science (19350030, 22350029)
and by Global COE Program of Kyoto University, "International Center for Integrated Research and Advanced Education in Materials Science",
from the Ministry of Education, Culture, Sports, Science and Technology of Japan.

\clearpage

\clearpage
\begin{table}
\centering
\caption{Parameters obtained from fitting the magnetic susceptibility to the isolated dimer\cite{dimer}
and the alternating chain\cite{Johnston} models.}
\begin{tabular}{|l|c|c|} \hline
 & Isolated dimer & Alternating chain \\ \hline
$J_0$ (K) & $26.7 \pm 0.3$ & $31.5 \pm 0.1$ \\
$J_1$ (K) & - & $23.0 \pm 0.9$ \\
g & $1.75 \pm 0.02$ & $2.09 \pm 0.01$ \\
$C$ (emu $\cdot$ K/mol) & $(6.78 \pm 0.47) \times 10^{-3}$ & $(5.86 \pm 0.13) \times 10^{-3}$ \\
$\chi_0$ (emu/mol) & $(6.04 \pm 0.61) \times 10^{-4}$ & $(-1.08 \pm 0.18) \times 10^{-4}$ \\ \hline
%fit & $J_1$ & $J_2$ & $g$ & $C$ & $x_0$ \\
% & K & K & &  $\times 10^{-3}$ emu K/mol & $\times 10^{-4}$ emu/mol \\ \hline
%1 & $31.5 \pm 0.1$ & $23.0 \pm 0.9$ & $2.09 \pm 0.01$ & $5.86 \pm 0.13$ & $-1.08 \pm 0.18$ \\
%2 & $26.7 \pm 0.3$ & - & $1.75 \pm 0.02$ & $6.78 \pm 0.47$ & $6.04 \pm 0.61$ \\ \hline
\end{tabular}
\label{parameters}
\end{table}

\end{document}